\documentclass{aa}

\usepackage{newtxtext,newtxmath}
\usepackage{amsmath,ulem}
\usepackage{physics}
\usepackage{multirow}
\usepackage{caption}
\usepackage{tablefootnote}

\usepackage[T1]{fontenc}
\DeclareRobustCommand{\VAN}[3]{#2}
\let\VANthebibliography\thebibliography
\def\thebibliography{\DeclareRobustCommand{\VAN}[3]{##3}\VANthebibliography}

\usepackage{dblfloatfix}
\usepackage{graphicx}	
\usepackage{amsmath}	
\usepackage{amssymb}	
\usepackage{hyperref}
\usepackage{xcolor}
\usepackage{float}

\title{Investigating the consistency of the shape and flux of X-ray reflection spectra in the hard state with an accretion disk reaching close to the black hole}

\author{
Sudeb Ranjan Datta \inst{1},
Michal Dov{\v{c}}iak \inst{1}, Michal Bursa \inst{1}, Wenda Zhang \inst{2}, Ji{\v{r}}{\'\i} Hor{\'a}k \inst{1}, Vladim{\'\i}r Karas \inst{1}
}

\institute{
Astronomical Institute of the Czech Academy of Sciences, Bo\v{c}n\'\i -II 1401, Praha 4, Prague, 141~00, Czech Republic\\
\email{datta@asu.cas.cz}
\and
National Astronomical Observatories, Chinese Academy of Sciences, 20A Datun Road, Beijing 100101, China\\
}

\titlerunning{Tension for reprocessed blackbody in hard state}  
\authorrunning{Datta et al.}

\date{Received September 15, 1996; accepted March 16, 1997}
\begin{document}

\abstract
{The observed spectra from black hole (BH) X-ray binaries (XRBs) typically consist of two primary components. A multitemperature blackbody originating from the accretion disk in the soft X-ray, and a power law-like component in the hard X-ray, due to the Comptonization of soft photons by the hot corona. The illumination of the disk by the corona gives rise to another key component known as reflection. A fraction of the incident hard X-ray radiation is naturally absorbed and re-emitted as a blackbody at lower energies and referred to as the ``reprocessed blackbody''.} 
{For densities relevant to XRBs and typical ionization values, the reprocessed blackbody may become significant in the soft X-ray region (approximately 0.1-1.0 keV) and should be noticeable in the observed spectra as a consequence of reflection. The absence of any blackbody component in the low/hard state of a BH XRB may not be consistent with the reflection of highly irradiating flux, observed as a power law from an appropriately dense disk of XRB.}
{We focus on the low/hard state of the BH XRB MAXI J1820+070. In contrast to previous works, we simultaneously fit the shape and flux of the reflection spectra. This allowed us to estimate the correct density and ionization of the slab as well as the corresponding reprocessed blackbody.}
{Our fitting of the representative observation of the BH XRB low/hard state suggests that the disk may, in principle, extend very close to the BH, even though the reprocessed thermal emission (due to disk illumination) remains cold (and thus low) enough to be consistent with the data in contrast to the results of a previous study. The inner reflection component is highly ionized and its fit is primarily driven by its contribution to the continuum, rather than by the shape of the relativistic iron line.}
{The reprocessed blackbody cannot help determine whether the disk extends close to the BH or not in the hard state. For this specific observation, the flux in inner reflection component turns out to be quite low with respect to the outer reflection or power law. The outflowing slab corona covering the inner region of the disk could be the plausible geometry of the source, with the underlying disk approaching near to the BH.}


\maketitle

\section{Introduction} 
\label{sec_intro}
Reflection spectra (RS) modeling in X-ray binaries (XRBs) or active galactic nuclei (AGNs) offers a promising approach to studying the space-time geometry around black holes (BHs) (\citealt{Miller2007, Fabian2010, Reynolds2014, Bambi2021}). To model the RS, it is typical to simulate the RS theoretically for a specific incident radiation illuminating a slab. Changes in incident radiation and/or the properties of the slab result in different final spectra, which are stored as tables known as reflection tables, such as {\tt reflionx} (\citealt{Ross2005, Tomsick2018}) and {\tt XILLVER} (\citealt{Garcia2010, Garcia2013, Garcia2016}). Very recently, similar reflection tables have been built including polarization as well \citep{Podgorny2022}. Theoretically simulated RS used to be convolved relativistically and integrated over the disk to fit the observed RS from an accretion disk (\citealt{Fabian1989, Laor1991, Dabrowski1997, Dovciak2004, Dauser2010, Garcia2014}). During the early development of reflection tables, the density ($n$) of the slab was fixed at $10^{15}$ cm$^{-3}$, which is the typical density for the accretion disk of AGNs. However, for XRBs, we expect that the density of the accretion disk, especially in the inner region, can be much higher than $10^{15}$ cm$^{-3}$ (\citealt{Frank2002book, Kato2008book}). Nonetheless, observations of XRBs are also typically fitted using reflection tables of $10^{15}$ cm$^{-3}$ with some arbitrary normalization factor to match the observed flux. The reasoning behind doing this is twofold. First, making reflection tables for high-density slabs is much more computationally and physically demanding (\citealt{Kallman2021}). With respect to RS, we are primarily concerned with the shape of the iron line, which is very much dependent on how much the slab is ionized, namely, the ionization ($\xi$) parameter of the reflection tables (\citealt{Ross2005, Ross2007, Garcia2010, Garcia2013, Tripathi2020}). Arbitrary normalization also hides the uncertainties in the distance to the source and its inclination. Thus, we can uncover the physical picture of the source to a great extent by focusing only on the shape of the observed spectra. However, it was recently reported that density can play a crucial role in altering the shape of the iron line, the Compton hump, and can provide thermalized blackbody emission in soft X-ray that is due to the reprocessing of the illumination (\citealt{Garcia2016, Tomsick2018}). The higher density disk in XRBs (compared to AGNs) makes the fitting of XRB data with high-density tables more reasonable.

\begin{figure*}
\centering\includegraphics[width=\textwidth]{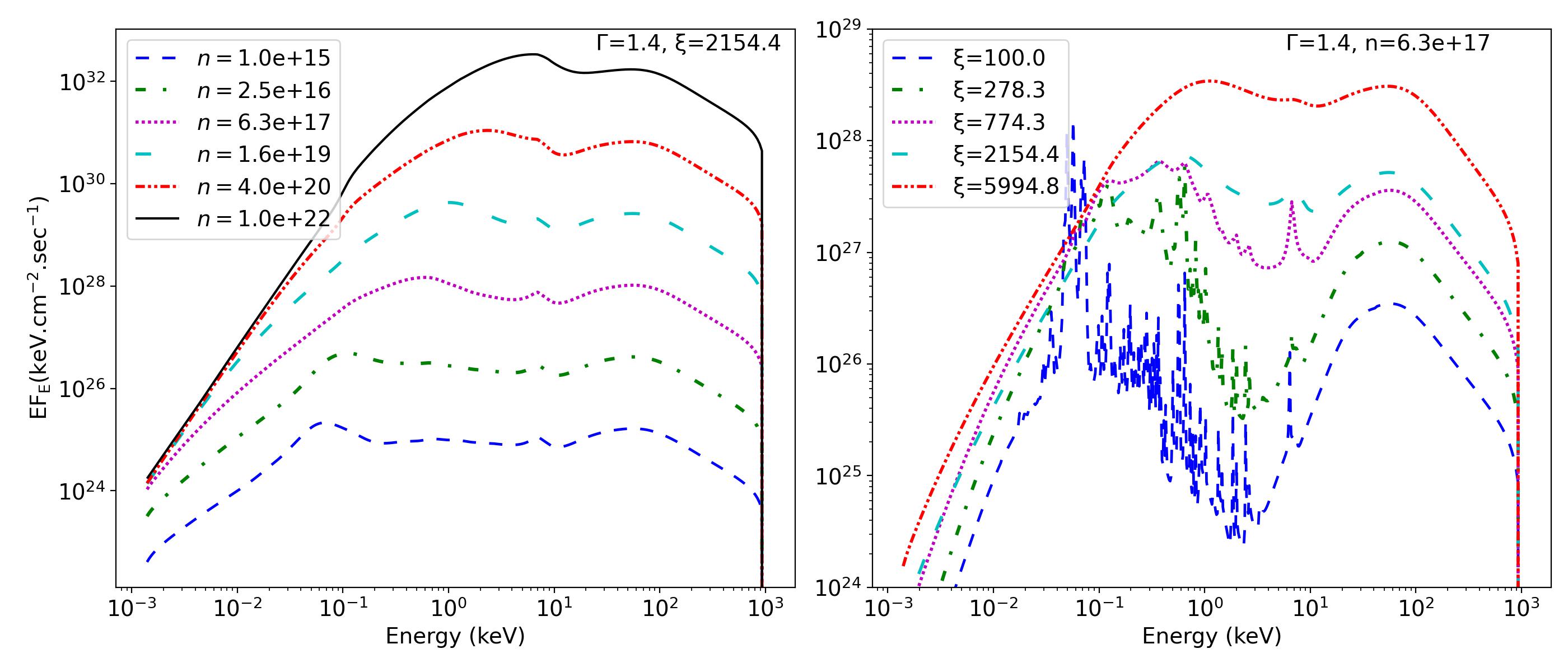}
\caption{Variation in EF$_{\rm E}$ with density ($n$) and ionization ($\xi$) from {\tt reflionx\_hd} table with appropriate normalization. The corresponding $n$ and $\xi$ are written in the legend. The {\tt reflionx\_hd} table has three input parameters: $n$, $\xi$, and $\Gamma$. For a variation of either $n$ or $\xi$, the fixed values of other input parameters are noted on the plot. More on the normalization of these spectra in footnote \ref{footnote_norm_reflionx_table}.}
\label{fig_reflionxhd_density_xi_variation}
\end{figure*}

As an essential consequence of the reflection, a fraction of incident radiation is always absorbed and emitted in lower energies. The albedo gives the fraction of incident radiation that is reflected back. From early studies, we know that the albedo of an ionized disk can vary in the range $\sim$ 0.3-0.7 (\citealt{Basko1974, Nayakshin2001, Psaltis2002}) depending on ionization, which means a significant fraction of the incident radiation will always be absorbed and emitted as blackbody and fluorescent lines. With the increment of the density of the illuminated slab, if we need the same $\xi$ to fit the RS, more illumination is required; this subsequently increases the thermalized blackbody radiation (i.e., the temperature at which the blackbody peak will appear). This is the reason for the appearance of thermalized blackbody radiation in the soft X-ray at higher slab densities, which is shown clearly in the left panel of Fig. \ref{fig_reflionxhd_density_xi_variation}. Due to the high dependence of the absorbed flux on illumination, blackbody radiation will also appear in soft X-rays for very high ionization cases, even at low density, as shown in the right panel of Fig. \ref{fig_reflionxhd_density_xi_variation}. If the observed RS demands sufficiently high ionization and density of the disk, then consequently the reprocessed blackbody will appear in soft X-ray. The effect of galactic absorption also sharply decreases from ultraviolet to X-ray. This makes it possible to observe the reprocessed blackbody component in soft X-ray if it is produced in the local frame of the source. However, in low/hard states of the XRBs, we do see RS but do not see any such blackbody radiation (\citealt{Remillard2006, Dunn2010}). This is the motivation behind our investigation of the consistency of the disk reflection of highly irradiating flux observed as a power law specifically in low/hard states of XRBs. One general shortcoming of using the {\tt XILLVER} or {\tt reflionx} tables is that none of them include intrinsic viscous dissipation. While this generates radiation in ultraviolet for AGNs and can be ignored for X-ray spectra, in the case of XRBs, this will certainly be significant (especially in high/soft state). The absence of any blackbody component in the low/hard state of the XRBs supports the usage of {\tt reflionx} or {\tt XILLVER}, which we present here. Although this issue has been addressed in {\tt refhiden} reflection table (\citealt{Ross2007}), the lower limit of the temperature of the blackbody due to viscous dissipation is 0.4 keV, making it suitable only in bright high/soft state.

A significant amount of work has been undertaken very recently about fitting XRB RS with high-density tables (\citealt{Tomsick2018, Jiang2019, Jiang2022, Chakraborty2021}, even in the hard state, \citealt{Liu2023}, for AGNs see \citealt{Jiang2019AGN, Mallick2022}). However, even with high density tables, the arbitrary normalization factor is always used to scale the flux as per the requirement. For reflection tables, $\xi$ and $n$ defines completely the illuminating flux ($F_{\rm inc}$). Again, RS is the radiation emitted by the slab when it is in radiative equilibrium with the incident radiation; this makes the reflected flux ($F_{\rm refl}$) equal to $F_{\rm inc}$ (\citealt{Garcia2010}) and makes it possible to define the reflected flux in the following way:
\begin{equation}
    F_{\rm refl}=F_{\rm inc}=\frac{\xi n}{4\pi}.
    \label{eqn_F_refl}
\end{equation}

Here, $F_{\rm refl}$ indicates the radiation emitted by the disk per unit area per unit time in its local frame. Unless stated otherwise, we always consider the full energy range 1 eV to 1 MeV to compute $F_{\rm refl}$, which also includes the reprocessed blackbody component. Therefore, once $\xi$ and $n$ are fixed, we know $F_{\rm refl}$. From the fitting of RS, not only do we get $\xi$ and $n$, but also the inner ($r_{\rm in}$) and outer ($r_{\rm out}$) radii of the disk (in units of gravitational radius), giving the area emitting $F_{\rm refl}$. For some XRBs, we know the distance, inclination, and mass of the BH quite well or at least the narrow range in which they lie (\citealt{Tetarenko2016}). Using the mass of the BH, we can find the gravitational radius and consequently the area of the disk in physical units. Using $F_{\rm refl}$ and total emitting area of the disk we can calculate the total luminosity emitted by the disk in its local frame as reflection. As the disk emits as a cosine source, from the distance and inclination, we can then estimate observed reflected flux ($F_{\rm refl,o}$) from the computed luminosity (Equation \ref{eqn_F_re_o}). These simple steps provide exact value of observed reflected flux from the fitted RS and takes away the requirement of arbitrary normalization. Therefore, if we know the mass of the BH, along with the distance and inclination of the source, it should be possible to fit the observed RS, considering both the shape and the flux. This may put a strong constraint on reflection in the low/hard state of XRBs due to reprocessed blackbody.


\cite{Zdziarski2020} (hereafter ZDM20) illustrated the concern over the absence of thermalized blackbody radiation in the hard state of XRBs by linking the primary radiation (the power-law) with the disk area. If the corona is on the axis and not very high from the disk, the illuminating flux at the innermost radius of the disk inversely depends (approximately) on the square of the radius (Equation (3) in ZDM20). If the disk is extended to the innermost stable circular orbit (ISCO), the illuminating flux at the innermost radius will be larger than in truncated disk geometry, creating more significant thermalized blackbody radiation due to reprocessing. For typical values of the albedo, the peak for thermalized blackbody should appear in the range 0.1-1.0 keV for the observed primary power law in the hard state of XRBs, which we do not observe (Fig. 3 of ZDM20, \citealt{Zdziarski2021}). In summary, the disk in the hard state of XRBs cannot approach too closely to the BH due to the restriction from thermalized blackbody as a consequence of illumination of the disk by primary hard X-ray. However, the above study requires the reflected flux to be connected with the primary power-law, which consequently includes the uncertainty in albedo as well as the geometry of the source. The amount of illumination (and thus also reflection) on the disk depends on the illuminating solid angle, namely, the geometry of the power-law emission region and its position with respect to the reflecting disk. Usually, the normalization of the reflection component is not linked with the normalization of primary power-law while fitting the observation due to geometrical uncertainty. The ratio between the two is typically denoted as reflection fraction \citep{Dauser2016}. Therefore, in this study, we also kept the normalization of the primary power-law free, independent of reflection component. However, as mentioned in the earlier paragraph, for RS, we made sure the fitting was consistent in terms of the shape and flux simultaneously, depending on the area of the disk, distance, and inclination to the source; with $\xi$ and $n$ define the amount of illumination and, thus, the reflection as well. With this setup, we can verify whether disk extending to ISCO or truncated geometry is better for high-density disk reflection incorporating self-consistent normalization of the RS simultaneously from its shape and flux.

Another pivotal issue is the emissivity profile, which requires attention in the context of this investigation. It is natural that due to a change in the illumination of the disk with radius, the reflection by the disk also changes. The change in emission of the disk with radii is represented through the emissivity profile: 
\begin{equation}
\label{eqn_emi_profile}
    I_\nu(r)\propto r^{-q},
\end{equation}
where $r$ is in units of $R_{\rm g}$ ($R_{\rm g}=\frac{GM}{c^2}$, gravitational radius) and $q$ is the emissivity index (\citealt{Wilkins2012, Dauser2013}). Also, $q$ plays a crucial role in fitting the broad Fe emission line present in the observed RS (\citealt{Svoboda2012, Fabian2014}). In lamp-post geometry, for large values of $r$ for the disk, compared to the height of the corona, the $q$ value will be 3. For the Shakura-Sunyaev disk, the $q$ value is also 3 (applicable for a radius much larger than zero torque inner boundary of the disk, Equation 3.52 in \citealt{Kato2008book}). This is relevant if the corona lies just above the disk, does not extend very close to the BH and is fed by the accretion process. We fix $q=3.0$ as done in ZDM20, \cite{Zdziarski2021MAXI, Marino2021} etc. Later in this work, we also assessed the robustness of our final conclusion on the assumed $q$ value.

A discussion of the source and observation is given in Section \ref{sec_formalism}. We describe how we modeled the RS in Section \ref{sec_modeling_ref_spectra}. First, we reproduce the expected inconsistency of reprocessed blackbody with observation in Section \ref{sec_problem}. In Section \ref{sec_kyn_model_and_fit} we build fully self-consistent model and report its results by simultaneously fitting the shape and flux of the RS including reprocessed blackbody. We discuss a few crucial points including possible geometry of the source in Section \ref{sec_discussion} and present our conclusions in Section \ref{sec_conclusions}. Another alternative model for fitting the shape and flux simultaneously is presented in Appendix \ref{appendix_density_iteration}. For completeness, we present the fitting results of {\it NuSTAR} and {\it NICER} observations in Appendix \ref{appendix_fit_nicer_nustar}.

\section{Observations and source details}
\label{sec_formalism}
We used {\it NuSTAR} observation of MAXI J1820+070 with ID 90401309002 during its low/hard state (Epoch 1 in \citealt{Zdziarski2021MAXI, Marino2021}). We reduced the observation using {\tt NuSTARDAS} following the standard procedure\footnote{https://heasarc.gsfc.nasa.gov/docs/nustar/analysis/} with heasoft version 6.33.1. The reasoning behind the choice of this observation is that our study is relevant in the low/hard state of XRBs when the thermal component is the weakest. In addition, following the detailed studies of this source (\citealt{Torres2019, Atri2020, Torres2020, Mikolajewska2022}), we fixed the BH mass to 8 $M_{\odot}$, distance to 3 kpc, inclination to 66$^{\circ}$, and hydrogen column density for absorption in inter stellar medium (N$_{\rm H}$ of the {\tt TBabs} model) to $1.4\times10^{21}$ cm$^{-2}$ (\citealt{Kajava2019, Zdziarski2021MAXI}, $1.5\times10^{21}$ cm$^{-2}$ is used in \citealt{Chakraborty2020}).

Although we used {\it NuSTAR} FPMA data within the energy range 3.0-78.0 keV for the fitting procedure, the effect of thermalization can be more prominent in the soft X-ray region below 3 keV. Therefore, we considered the quasi simultaneous observation by {\it NICER} with ID 1200120104 for energy range 0.5-10.0 keV. We reduced the {\it NICER} data using {\tt NICERDAS} software version 2024-02-09\_V012 in heasoft 6.33.1, following the standard procedure. We used the latest {\it NICER} calibration files of CALDB release ``xti20240206'' to extract level 2\footnote{https://heasarc.gsfc.nasa.gov/docs/nicer/analysis\_threads/nicerl2/} data and, finally, level 3 spectral data, rmf, and arf files\footnote{https://heasarc.gsfc.nasa.gov/docs/nicer/analysis\_threads/nicerl3-spect/} (which include a 1.5\% spectral systematic error in the energy range 0.5-10.0 keV). However, the low-energy data from {\it NICER} typically reveal some calibration features and require some phenomenological models (e.g., {\tt gabs, edge}) for the fitting procedure \citep{Wang2020, Wang2021, Ludlam2021, Ludlam2022, Kumar2022, Feng2022, Moutard2023, Svoboda2024ApJ, Svoboda2024ApJSwiftSoft}. In addition, we have to take into account the constant factors for cross-normalization across different instruments to fit simultaneous data from {\it NuSTAR} and {\it NICER}. Therefore, to keep the fitting procedure simple and to concentrate mainly on the reprocessed blackbody due to reflection, we fit the {\it NuSTAR} FPMA data only in the energy range of 3.0-78.0 keV; we show the {\it NICER} data in the plot for energy range of 0.5-10.0 keV. This efficiently demonstrates whether the fitted model for {\it NuSTAR} data is overestimating the reprocessed blackbody (or not) with respect to the simultaneous low-energy data from {\it NICER}. If the model flux remains below the {\it NICER} data, then we can safely say that reprocessed blackbody due to illumination of the disk does not contradict the non-observation of any blackbody component in the hard state. In Appendix \ref{appendix_fit_nicer_nustar}, we present the complete fitting of {\it NICER}, {\it NuSTAR} FPMA and {\it NuSTAR} FPMB data with our fully self-consistent model (details in Section \ref{sec_kynrefionx_model}) which 
fits the shape and flux of the RS simultaneously including reprocessed blackbody. We do not focus on the spin measurement of the BH here, instead freezing it to 0.998 for the entire study.

\section{Modeling the RS}
\label{sec_modeling_ref_spectra}
From the existing literature, it is rather clear that one reflection component is insufficient to fit the observed RS for most of the sources (\citealt{Wang2018, Sridhar2020, Liu2023}). For the observations of MAXI J1820+070 as well, the inclusion of two reflection components (a broad component for relativistic reflection and a narrow component for distant reflection) improves the spectral fits significantly (\citealt{Buisson2019, Chakraborty2020, Zdziarski2021MAXI, Marino2021}). Therefore, we also include two reflection components to fit the RS. To model the interstellar absorption and primary power law, we used {\tt TBabs} and {\tt cutoffpl}, respectively, in {\tt XSPEC}\footnote{https://heasarc.gsfc.nasa.gov/xanadu/xspec/} \citep{Arnaud1996}. Therefore, our complete model for fitting becomes {\tt TBabs}$\times$({\tt cutoffpl} + reflection + reflection). To model the reflection, we have two choices available, which are mostly used by the community: {\tt reflionx} and {\tt XILLVER}. However, publicly available version of {\tt XILLVER} tables\footnote{https://sites.srl.caltech.edu/~javier/xillver/} only cover the energy range from 0.1 keV - 1 MeV; whereas our work (specifically model C in Appendix \ref{appendix_density_iteration}) incorporates the reprocessed blackbody demands estimation of flux in broader energy range of 1 eV - 1 MeV. The flux estimation in this broad energy range also plays a crucial role in improving our understanding of the geometry of the source (see Section \ref{sec_diss_geometry}). For this reason, we used {\tt reflionx\_hd} table\footnote{reflionx\_HD.mod, available at https://www.michaelparker.space/reflionx-models} to model the high-density disk reflection. This table is prepared by illuminating the slab with a hard X-ray modeled by {\tt cutoffpl} and fixing the cutoff energy at 300 keV (\citealt{Tomsick2018, Ross2005}). Here we fit the primary hard X-ray with {\tt cutoffpl} setting the cutoff energy at 300 keV, and the photon index of the primary is linked to the photon index of the {\tt reflionx\_hd} table, which ensures the consistency between primary and reflection. Also, the {\tt reflionx\_hd} table assumes the iron abundance to be solar. The final models differ, depending on how the reflection components are modeled.

At first, to reproduce the expected discrepancy of observation with reprocessed blackbody, we fit the shape of the reflection component, assuming the disk density to be $10^{15}$ cm$^{-3}$, and estimated the reprocessed blackbody from the reflection (see Section \ref{sec_problem}). Next, we built a model with high-density disk reflection that includes reprocessed blackbody in the fitting and self-consistent methodology to fit the observation simultaneously by its shape and flux, as described in Section \ref{sec_kyn_model_and_fit}.


\begin{figure}
\centering\includegraphics[width=\columnwidth]{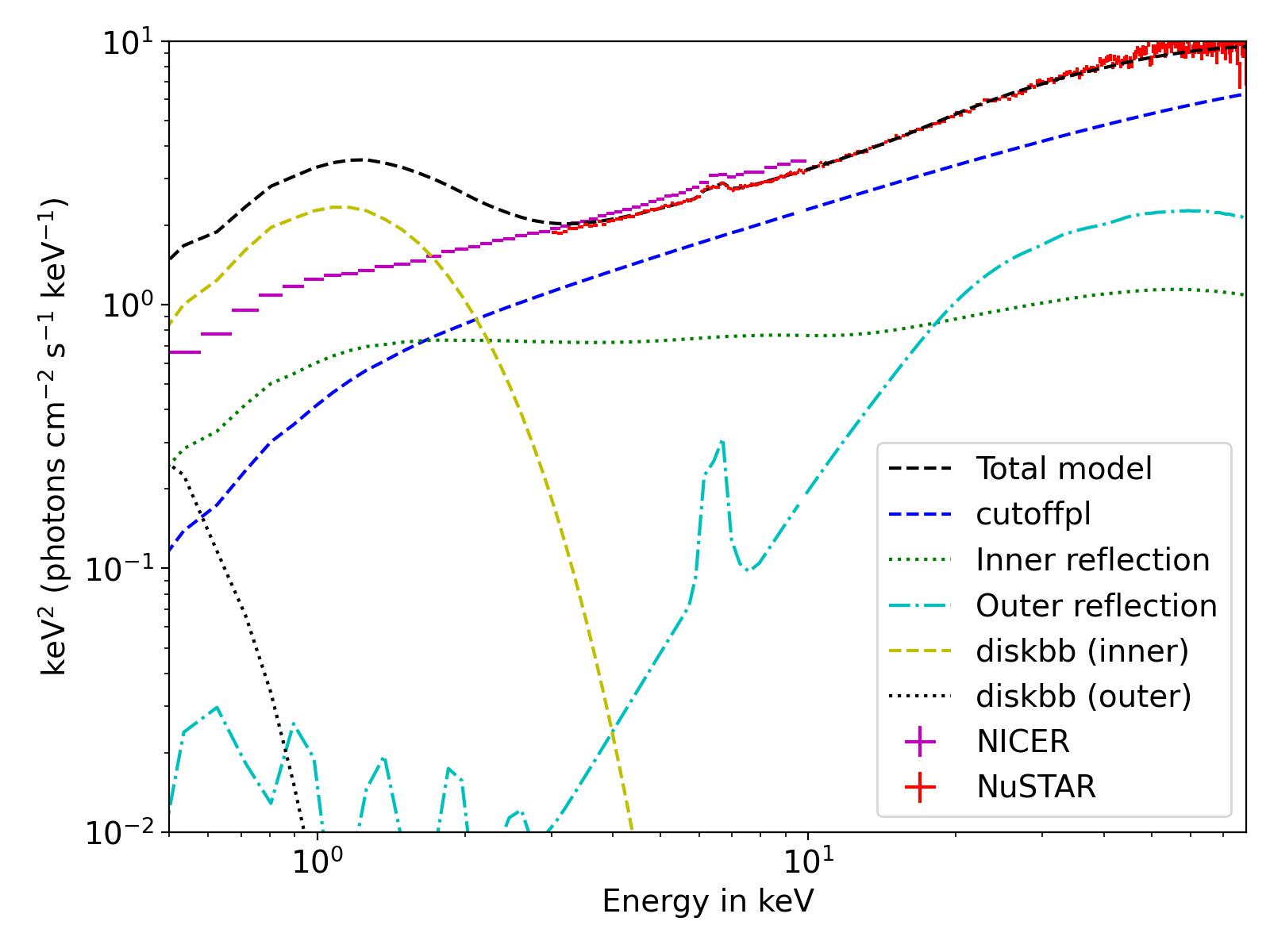}
\caption{Best-fit model A for {\it NuSTAR} data. Expected reprocessed blackbody as a consequence of reflection are modeled as {\tt diskbb} components, which are added separately over the best-fitted model A. {\tt diskbb}(inner) and {\tt diskbb}(outer) denote the {\tt diskbb} components corresponding to thermalization due to inner and outer reflection, respectively. Simultaneous {\it NICER} data were presented to compare the fitted model in the lower energy. We present the unfolded spectrum to show the data of different instruments on the same scale.}
\label{fig_the_problem}
\end{figure}

\subsection{Inconsistency with the reprocessed blackbody}
\label{sec_problem}
To fit the reflection component, it is typical to convolve (using {\tt kyconv}: \citealt{Dovciak2004}, or {\tt relconv}: \citealt{Dauser2010, Garcia2014}) the reflection spectrum of the {\tt reflionx\_hd} table to include the general relativistic effects and the contribution from the whole disk. To understand the problem of reprocessed blackbody better and to reproduce it, we first followed ZDM20 to a great extent (but not fully) to fit the observed spectra by treating the reprocessed blackbody separately from the reflection component. This is achieved by fixing the density of the slab for reflection to be $10^{15}$ cm$^{-3}$, making our first model,
\newline
model A = {\tt TBabs}$\times$({\tt cutoffpl} + {\tt relconv}$\otimes${\tt reflionx\_hd} ($10^{15}$ cm$^{-3}$)+ {\tt relconv}$\otimes${\tt reflionx\_hd} ($10^{15}$ cm$^{-3}$)).
\newline
Here we kept the normalization parameter of the {\tt reflionx\_hd} table in {\tt XSPEC} free while fitting. The best fit with this model ($\chi^2/d.o.f=2157/1867$) gives a reasonable estimate of the reflection. To estimate the unabsorbed reflected flux, we computed the flux corresponding to reflection component within the energy range of 0.1-1000.0 keV (different from 1 eV - 1 MeV range used in rest of the paper) to ignore any possible contribution from thermalization (\citealt{Dovciak2022}). We assumed that there is no reprocessed blackbody radiation above 0.1 keV for density of $10^{15}$ cm$^{-3}$ (same as in ZDM20). Although the boundary of energy between reflection and reprocessed blackbody is vague, this is the best we can do if we treat those separately. We note that the RS in {\tt reflionx\_hd} table includes the reprocessed blackbody radiation within full energy range, but by fixing the density to $10^{15}$ cm$^{-3}$ and lower boundary of energy to 0.1 keV, we may be able to ignore the reprocessed blackbody from the reflection. Assuming an average value of albedo of 0.5, we can consider that the flux emitted as reflection should give rise to similar blackbody flux, which we add as a separate component using the model {\tt diskbb} (\citealt{Mitsuda1984}). From the fitted inner radius of the disk ($r_{\rm in}$), and using distance ($D$) and inclination ($\theta$) of the source, we can easily fix the normalization ($A_{\rm dbb}$) of {\tt diskbb} component, given by: 
\begin{equation}
A_{\rm dbb}=\left(\frac{r_{\rm in}/km}{D/(10 kpc)}\right)^2\cos\theta.
\end{equation}
We can then find the temperature that equates the {\tt diskbb} flux to reflected flux (Equation (10) in \citealt{Zdziarski2021}). We note that here $r_{\rm in}/km$ indicates the inner radius of the disk in units of km. Flux in 0.1 keV - 1000 keV energy range for the inner and outer reflection component becomes $1.0\times10^{-8}$ and $8.1\times10^{-9}$ ergs.cm$^{-2}$.sec$^{-1}$, respectively. Also, the fitting gives $r_{\rm in}$ to be 7.5 R$_{\rm g}$ and 167.5 R$_{\rm g}$ which set the T$_{\rm in}$ of the added {\tt diskbb} to 0.35 keV and 0.08 keV, respectively, for the inner and outer reflection component (using $D=3$ kpc, $\theta=66^\circ$). The crucial difference between the above procedure and what followed in ZDM20 is that here we equate the blackbody flux with reflected flux instead of connecting it with the primary\footnote{In ZDM20, the luminosity of the primary power-law component is estimated at first assuming spherically symmetric corona, and then illuminated flux on the disk is computed depending on the inner radius of the disk.}.

Figure \ref{fig_the_problem} shows that the reprocessed blackbody ({\tt diskbb} component) due to inner reflection overshoots the data in the soft X-ray region significantly. This raises concerns because the reprocessed blackbody component must arise as a consequence of reflection, although we never observe any blackbody component along with the reflection component in the hard state of XRBs (as pointed out in ZDM20). For the outer reflection component, a high value for $r_{\rm in}$ makes the corresponding {\tt diskbb} normalization large. Therefore, to achieve similar flux, $T_{\rm in}$ for the outer reflection becomes much lower compared to the inner reflection. The above exercise shows that despite the inclusion of two reflection components instead of one, and even estimating the reprocessed blackbody from reflected flux instead of connecting it with primary radiation, the inner reflection still leads to strong reprocessed soft X-ray emission, which is not present in the observed data. However, two crucial points are overlooked here. We added the reprocessed blackbody components on top of the already fitted model, which makes the result not robust. Ideally, the fitting should be done by including the reprocessed blackbody components because multiple components can eventually be adjusted among themselves and ultimately result in a consistent fit. Treating the reprocessed blackbody separately from reflection is another tricky undertaking, due to the ambiguous boundary between the two, which is very sensitive to the density and ionization of the slab (Fig. \ref{fig_reflionxhd_density_xi_variation}). Therefore, to draw any inference about the thermalization, we chose more accurate models in the next section (Section \ref{sec_kyn_model_and_fit}) to fit the observations.

\subsection{Fitting the shape and flux of RS simultaneously including reprocessed blackbody}
\label{sec_kyn_model_and_fit}

As explained in Section \ref{sec_intro} (and later in more detail), from the fitting of shape of the RS emitted by the accretion disk, we can get all the required information to find the observed reflection flux. So, we should be able to connect the shape of the RS with its flux (i.e., photon counts) and avoid the arbitrary normalization in fitting. On the other hand, in the full energy range of 1 eV - 1 MeV, RS from {\tt reflionx\_hd} includes both the reflection (i.e., Compton hump, emission lines) and the reprocessed blackbody. Therefore, the best model to infer the reprocessed blackbody would fit the shape and flux of the RS together, including the reprocessed blackbody. This is achieved completely by incorporating {\tt reflionx\_hd} within the {\tt kynrefionx} model, including all the general relativistic effects (as detailed in Section \ref{sec_kynrefionx_model}).

\subsubsection{Modeling RS with kynrefionx}
\label{sec_kynrefionx_model}

\begin{figure}
\centering\includegraphics[width=\columnwidth]{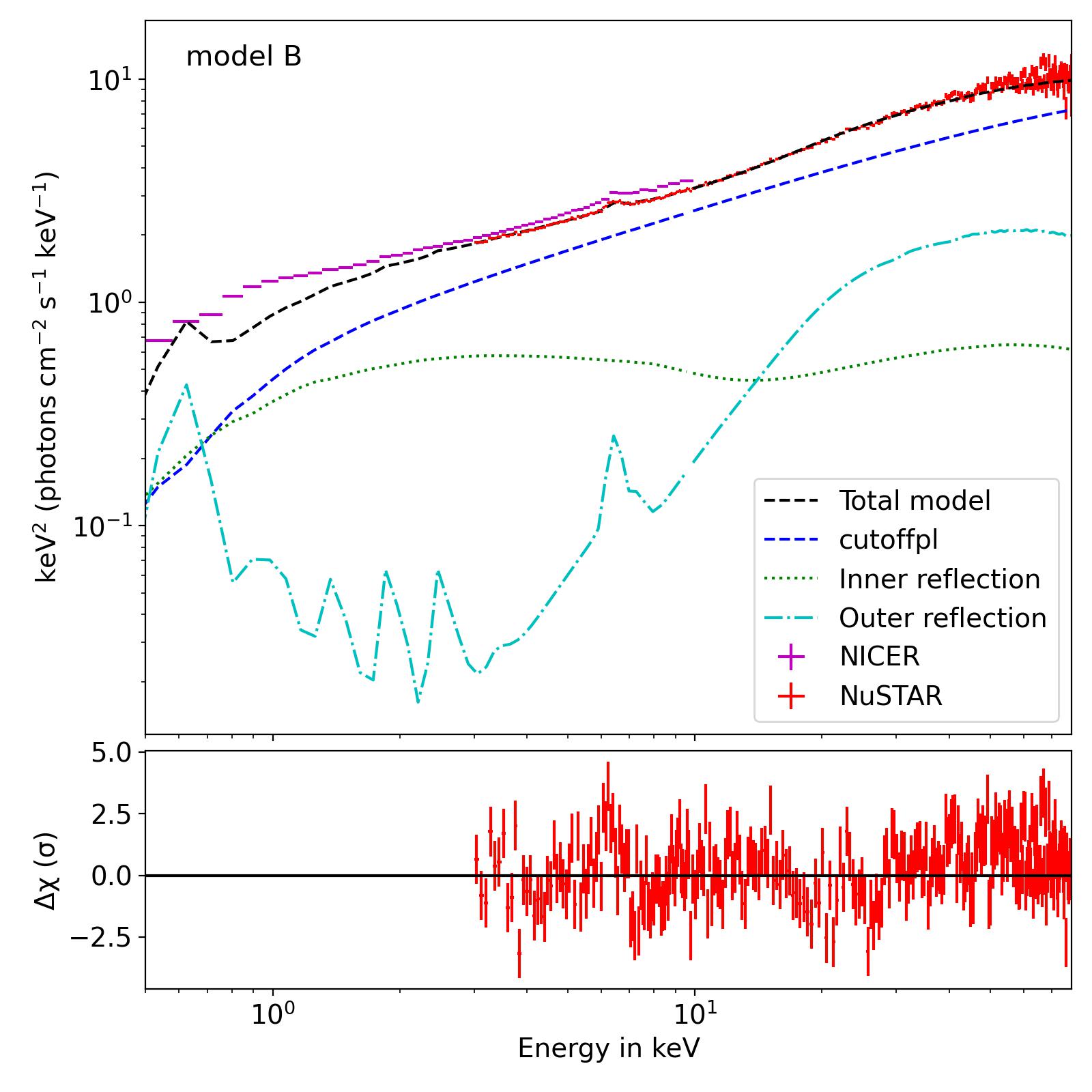}
\caption{Best-fit model B for the {\it NuSTAR} observation with reflection $\equiv$ {\tt kynrefionx}({\tt reflionx\_hd}). The fitted model parameters are presented in Table \ref{table_fit_results}. Simultaneous {\it NICER} observation is also overplotted in the energy range of 0.5-10.0 keV which is sufficient to understand the consistency or inconsistency of reprocessed blackbody. The primary reason for not fitting {\it NICER} data is the requirement of more model components (e.g., {\tt gabs}, {\tt edge}) to fit the different calibration features of the instrument at low energy which will make the model unnecessarily complicated. For completeness, in Appendix \ref{appendix_fit_nicer_nustar} we fit the {\it NICER} data as well with {\it NuSTAR}. Please note that we are showing unfolded spectra instead of actual data to match the Y-axis for two different instruments. Data were rebinned using the command ``setplot rebin 100 10'' to present in the plot.}

\label{fig_bestfit_kynrefionx}
\end{figure}


\cite{Dovciak2004} developed a fitting scheme within {\tt XSPEC} to study the X-ray spectra in strong gravity regime. It gives us flexibility to compute the spectra emitted by an accretion disk because of the opportunity to include the local emission of the disk separately within the code. Once the local emission (emission by the disk in its co-rotating local frame) at different radii of the disk are defined, the {\tt KY} code integrates it over the disk incorporating the general relativistic effects on radiation in Kerr space-time. There are different versions\footnote{https://projects.asu.cas.cz/stronggravity/kyn} of the code to model different features of the accretion disk as outlined in Table 2 of \cite{Dovciak2004}. One such version is {\tt kynrefionx}, which specifically can read {\tt reflionx} tables and fit the observed RS, while including or excluding the primary. We refer to \cite{Dovciak2004Thesis} for more details. 

We modeled the RS independent of primary power-law and fix the emissivity index to 3.0, as done for model A in the previous section. We incorporated the {\tt reflionx\_hd} table as the local emissivity of the disk within {\tt kynrefionx}. We assumed that the density of the disk remains same with radius and only $\xi$ (and consequently flux) changes with radius depending on the value of the emissivity index. The code basically takes spectra corresponding to different $\xi$ from the reflection table for different radii of the disk and integrates them over the disk area, including all general relativistic effects to give the final observed spectra. Due to the inclusion of RS of different ionization at different radii of the disk, this model is more accurate compared to model A, where the convolution model is applied externally; thus, it takes only one spectrum to convolve corresponding to one $\xi$ value for all radii. Another crucial step forward is the exact normalization of the spectra. In {\tt kynrefionx}, local emission of disk for reflection at a specific radius is taken from the {\tt reflionx\_hd} table, depending on $\xi$, $n$ at that radius and normalized such that the total reflected flux in the energy range of 1 eV - 1 MeV becomes equal to ($\xi n/4\pi$), following Equation \ref{eqn_F_refl}\footnote{Note: {\tt reflionx\_hd}, similarly to other available reflection tables, provides RS in some arbitrary units. It is necessary to contact the author of the table to get the normalization constant to find the spectrum in physical units. A possible workaround is to normalize the RS by assuming it is in radiative equilibrium with the incident radiation ($=\xi n/4\pi$). Here, we normalize all the spectra in {\tt reflionx\_hd} table such that total flux in full energy range 1 eV to 1 MeV becomes $=\xi n/4\pi$ to find the physical units and incorporate them in {\tt kynrefionx}. A few RS of {\tt reflionx\_hd} table with a proper normalization are plotted in Fig. \ref{fig_reflionxhd_density_xi_variation}.\label{footnote_norm_reflionx_table}}. As the local emission from each radius is normalized appropriately, the {\tt kynrefionx} code integrates these local reflected spectra throughout the whole disk from $r_{\rm in}$ to $r_{\rm out}$, which basically gives the reflected luminosity as a function of energy. To find the RS (and flux) in observer's frame, the code then renormalizes the spectra by the factor $(\cos\theta/\pi D^2)$, as done in Equation \ref{eqn_F_re_o}, and also modifies the spectra according to all general relativistic effects. The normalization parameter ($=(Mpc/D)^2=1.11\times10^5$ for $D=3$ kpc) in the {\tt kynrefionx} model incorporates distance to the source and inclination is provided in the model as another input parameter, which we fixed to 66$^\circ$. The model spectra produced by the above procedure are then used by {\tt XSPEC} to fit the observed RS. The above modeling procedure enables us to fit the flux of the observed RS along with its shape because the reflection spectrum itself (from {\tt kynrefionx}) is normalized appropriately and removes all arbitrariness in the normalization. All of these advantages lead to our final model:
\newline
model B = {\tt TBabs}$\times$({\tt cutoffpl} + {\tt kynrefionx}({\tt reflionx\_hd}) + {\tt kynrefionx}({\tt reflionx\_hd})).
\newline
The best-fit model B gives the density and ionization of reflection modeling, which are consistent in terms of both the flux and the shape, including all the general relativistic effects and radial variation of ionization.

It is also possible to fit both the shape and flux of the spectra keeping the model the same as in model A (i.e. reflection $\equiv$ {\tt relconv}$\otimes${\tt reflionx\_hd}), but extending the energy range from 1 eV to 1 MeV, along with the iteration of the fitted density and ionization parameter. Although this procedure is not as accurate as that of model B, it provides more flexibility to extend the same procedure to different reflection tables (if available) in the full energy range from 1 eV to 1 MeV. The method of iteration to find a consistent fit with this model and its fitting results, which are very similar to the results with model B are detailed in Appendix \ref{appendix_density_iteration}.


\begin{table*}
	\centering
	\caption{Best-fit model parameters with consistent flux and shape for model B and B.1.}
	\label{table_fit_results}
	\begin{tabular}{cccc}
		\hline
		Component & Parameter & Model B & Model B.1\\ 
		\hline
		Interstellar absorption & N$_{\rm H}$ (10$^{22}$ cm$^{-2}$) & 0.14(f) & 0.14(f)\\
		\hline
		   & $\Gamma$ & 1.385$_{-0.005}^{+0.009}$ & 1.371$_{-0.006}^{+0.004}$\\  
         Power law & cut-off(keV) & 300(f) & 300(f)\\
         & Norm & 0.649$_{-0.009}^{+0.007}$ & 0.622$_{-0.015}^{+0.013}$\\
         & Flux & 4.97$\times10^{-8}$ & 5.06$\times10^{-8}$\\
         \hline
          & $r_{\rm in}$ ($R_{\rm g}$) & 2.11$_{-0.15}^{+0.23}$ & 1.79$_{-0.04}^{+0.58}$\\
          & $r_{\rm out}$ ($R_{\rm g}$) & 1000(f)$^a$ & 1000(f)$^a$\\
          & $L$ (L$_{\rm Edd}$) & 2.81$_{-0.07}^{+0.09}\times10^{-4}$ & 2.52$_{-0.03}^{+0.04}\times10^{-3}$\\
          Inner & $\xi_{\rm in}$ & 1.06$\times10^{5}$ & 1.14$\times10^{6}$\\
          reflection& $n$ ($10^{15}$ cm$^{-3}$) & 1.24$_{-0.29}^{+0.32}\times10^{4}$ & 1.49$_{-0.25}^{+0.46}\times10^{4}$\\
          & Norm & 1.11$\times10^5$(f) & 1.11$\times10^4$(f)\\
          & Flux & 6.24$\times10^{-9}$ & 5.36$\times10^{-9}$ \\
          \hline
          & $r_{\rm in}$ ($R_{\rm g}$) & 671$_{-74}^{+71}$ & 669$_{-78}^{+76}$\\
           & $r_{\rm out}$ ($R_{\rm g}$) & 1000(f) & 1000(f)\\
           & $L$ (L$_{\rm Edd}$) & 1.18$_{-0.02}^{+0.02}\times10^{-3}$ & 1.14$_{-0.03}^{+0.01}\times10^{-3}$\\
          Outer & $\xi_{\rm in}$ & 107 & 107\\
          reflection & $n$ ($10^{15}$ cm$^{-3}$) & 1.55$_{-0.10}^{+0.11}\times10^3$ & 1.55$_{-0.10}^{+0.13}\times10^3$\\
          & Norm & 1.11$\times10^5$(f) & 1.11$\times10^5$(f)\\
          & Flux & 2.58$\times10^{-8}$ & 2.60$\times10^{-8}$\\
          \hline
           & $\chi^2/\nu$ & 2158/1866 & 2146/1866\\
          \hline          
	\end{tabular}
\flushleft
\textit{Notes.} {In model B, the reflection is modeled by {\tt kynrefionx(reflionx\_hd)} and its normalization parameter is fixed according to the distance of the source. Model B.1 is model B with modification in the normalization of inner reflection, introduced in the last paragraph of Section \ref{sec_diss_geometry}. Given errors represent 1 $\sigma$ range of fitted parameters. Luminosity ($L$) is a free parameter within {\tt kynrefionx}, representing the luminosity in units of Eddington luminosity (L$_{\rm Edd}$) from the extended corona at inclination 66$^\circ$ in 2-10 keV energy range illuminating the disk to give rise to the RS. $\xi(r)$ is computed inside the model and its value at the inner radius ($\xi_{\rm in}$) is presented. The tabulated flux is the unabsorbed one in units of ergs.cm$^{-2}$.sec$^{-1}$, calculated from {\tt XSPEC} within energy range 1 eV - 1 MeV. The frozen parameters during fitting are indicated as (f). We assume emissivity index $q=3$ and no radial variation of density. However, keeping $q$ to be free or assuming the radial variation of density to be the same as that of the Keplerian disk leads to very similar best fitted parameters.

$^a$Setting $r_{\rm out}$ of inner reflection = $r_{\rm in}$ of outer reflection leads to very similar best fitted results.}
\end{table*}

\subsubsection{Results}
\label{sec_kynrefionx_result}
Figure \ref{fig_bestfit_kynrefionx} and Table \ref{table_fit_results} give the best-fit results of {\it NuSTAR} FPMA data with model B. The presence of a reprocessed blackbody hump in the soft X-ray region (especially for the inner reflection) can be noticed easily due to the higher density value. To find the consistency of reprocessed blackbody with observed data, a simultaneous {\it NICER} observation in 0.5-10.0 keV is also very helpful; its unfolded spectrum is shown in Fig. \ref{fig_bestfit_kynrefionx}. Even then, the reprocessed blackbody radiation does not lead to any problems; that is, the model prediction is below the observed data, as opposed to what is shown in Fig. \ref{fig_the_problem} of current work or in Fig. 3 of ZDM20. This result implies that using the high-density tables, we can fit the low/hard state of MAXI J1820+070 self-consistently. Although the model fits the energy range below 40 keV very well, some residual remains in the higher energy. However, the significance of the data above 40 keV is low and possibly other adjustments might improve the high energy fitting (\citealt{Zdziarski2021MAXI, Marino2021}); namely, using {\tt nthcomp} to model the primary hard X-ray or making the cutoff energy free in the existing {\tt cutoffpl} model. The most crucial point to notice is the inner radius of the disk ($r_{\rm in}$), which turns out to be 2.1 $R_{\rm g}$. This exercise concludes that even within the curved space-time regime, the high-density disk can reach very near to ISCO, remaining consistent with the observed flux and shape simultaneously in low/hard state of XRBs. As the observed spectra is very similar to the typical spectra in the low/hard state of XRBs, we can generally say that it is most likely the reprocessed blackbody due to disk reflection is consistent, even if the disk extends close to the BH. We note that the compatibility of simultaneous {\it NICER} observation with the model spectra (from fitting only {\it NUSTAR} FPMA) in the lower energy in Fig. \ref{fig_bestfit_kynrefionx} is sufficient to draw a conclusion on the consistency of the reprocessed blackbody with the reflection. The primary reason for not fitting {\it NICER} data is the requirement of more model components (e.g., {\tt gabs}, {\tt edge}) to fit the different instrument's calibration features at low energy, which would make the modeling unnecessarily complicated. For completeness, in Appendix \ref{appendix_fit_nicer_nustar}, we fit the {\it NICER} data as well along with {\it NuSTAR} FPMA and FPMB.

As the disk approaches closer to the BH, fixing emissivity index to $q=3$ may not be a good assumption for fitting the inner reflection component (\citealt{Svoboda2012, Dauser2013}). If we keep $q$ for the inner reflection free, then the fit is improved slightly and the best fit gives $q=6.6$, which is more acceptable ($\chi^2/d.o.f=2145/1865$). We also tried more realistic configuration by assuming radial variation of density of the disk same as Keplerian disk. We assumed the density of the disk for inner and outer reflection to vary as $r^{-1.65}$ and $r^{-1.875}$, respectively, following the middle and outer region of the Keplerian thin disk (Equations 3.66 and 3.68 in \citealt{Kato2008book}). This imposition also improves the fitting slightly ($\chi^2/d.o.f=2134/1866$). However, with all these  sophisticated adjustments, the fitting gives $r_{\rm in}\sim 2-2.5 R_{\rm g}$, keeping the flux distribution in different components same as for model B. Even if we assume the BH to be slowly spinning ($a=0.5$), the fitted $r_{\rm in}$ becomes the same as the ISCO (4.23 R$_{\rm g}$ for $a=0.5$), making the fit slightly worse ($\chi^2/d.o.f=2167/1866$). Considering the flux contribution from different components, model B.1 (i.e., model B with a modification in the normalization of the inner reflection) is introduced (see Section \ref{sec_model_C1}) to make the geometry physically acceptable. Also with respect to model B.1, the disk extends very close to the BH, similarly to the results we see from model B.

\section{Discussion}
\label{sec_discussion}
\subsection{Reasons behind achieving a consistent good fit}
A higher value for $r_{\rm in}$ typically (unless it is very close to $r_{\rm out}$) requires the density and ionization to be lower to reach the same flux, if the other parameters remain the same (Equation \ref{eqn_F_re_o_limit}). This is why a greater flux in the outer reflection component still requires lower density and ionization than the inner reflection (Table \ref{table_fit_results}). This is exactly in line with the logic put forward by ZDM20, and higher $r_{\rm in}$ values put the thermal blackbody radiation at lower energy. It must be emphasized that if we treat the reflection and reprocessed blackbody separately, and then add the resulting blackbody as a consequence of reflection on top of an already fitted model (Section \ref{sec_problem}), then it is very probable that model will overshoot the observed spectra (Fig. \ref{fig_the_problem} in this paper, Fig. 3 in ZDM20), making the reprocessed blackbody inconsistent with the observed reflection. On the contrary, fitting the shape and flux simultaneously, while including the reprocessed blackbody (Section \ref{sec_kyn_model_and_fit}), results in very good fit (Fig. \ref{fig_bestfit_kynrefionx}, Table \ref{table_fit_results}) even if the disk extends very close to ISCO. Another reason for the appearance of thermal hump in ZDM20 is to the connection between primary and reflection, which we typically ignore when we fit the spectra, as done in this study due to uncertainty in the geometry of the corona and its orientation with respect to the disk. The best-fit result with model B suggests that the inner reflection component is heavily ionized (Fig. \ref{fig_bestfit_kynrefionx}, Table \ref{table_fit_results}) and practically does not give any Fe line feature. Nevertheless, it plays a key role in the fit by providing some curvature near the Fe line and in the overall continuum. It is evident that if the shape of the spectra demands high $\xi$ and $n$, then naturally $r_{\rm in}$ has to be low enough to match the flux requirement (Equation \ref{eqn_F_re_o_limit}). The perfect balance between $\xi$, $n$, and $r_{\rm in}$ gives a fitting result that consistent in terms of both the shape and the flux.

\begin{figure*}
\centering\includegraphics[width=\textwidth]{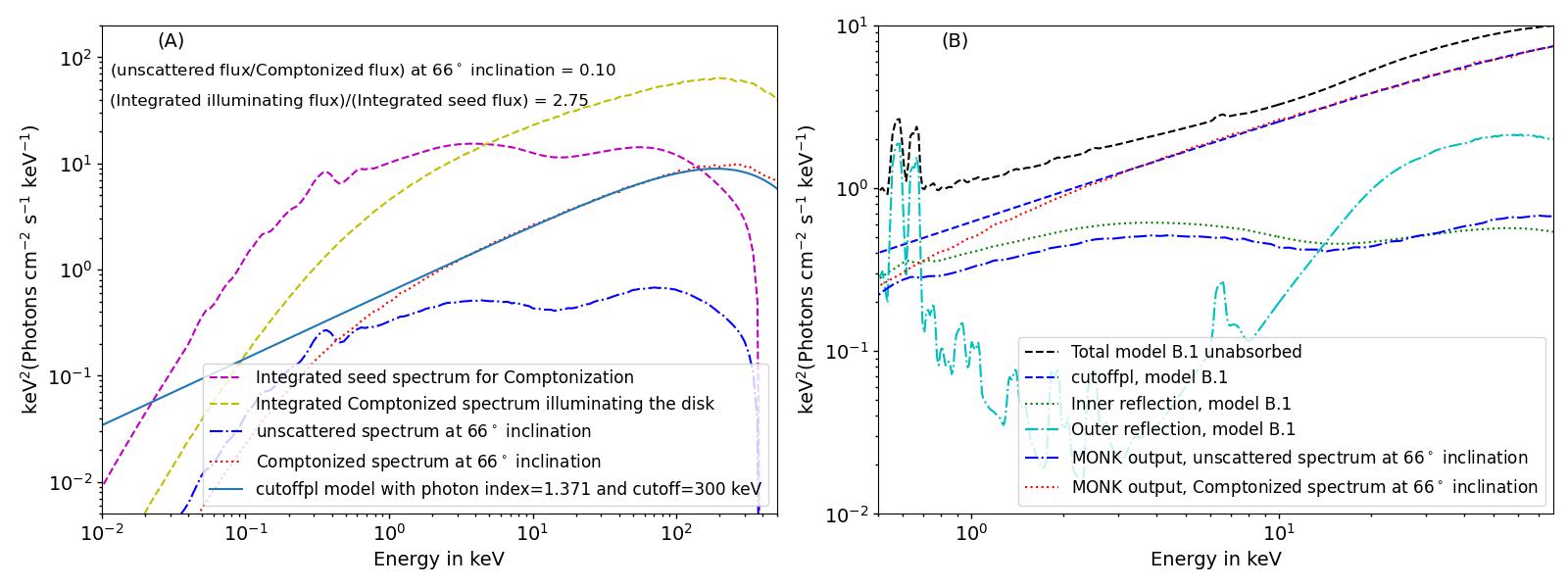}
\caption{Comptonization of inner reflection photons of model B.1 through slab corona of optical depth 0.282 and electron temperature 160 keV using MONK. Panel (A) shows the broadband spectral output from MONK. Integrated term is used to denote total spectrum integrated over 2$\pi$ solid angle, whereas 66$^\circ$ inclination is mentioned to indicate observation at that specific inclination. Seed spectrum is provided as the seed photons for Comptonization in MONK. Comptonized spectrum is made of photons which are scattered one or more times by the hot electrons in the corona. Unscattered spectrum is made of photons which are able to leak through corona without scattering. Flux is always computed in energy range 1 eV - 1 MeV. Panel (B) shows the unscattered spectrum and Comptonized spectrum of MONK output at inclination 66$^\circ$ along with model components of the best fitted model B.1 presented in Table \ref{table_fit_results}.}
\label{fig_monk_with_modelC1}
\end{figure*}

\subsection{Geometry of the source}
\label{sec_diss_geometry}
Although our best fit suggests that the disk is extended very close to the BH, making the spectral fits consistent from both the shape and the flux, it does not tell us anything about the geometry of the source. The flux present in different spectral components can guide us in understanding this aspect. Due to the large energy band (1 eV - 1 MeV), the flux represents the total radiation available in respective components. From the fitting results of model B (Table \ref{table_fit_results}), we find that the total flux in inner and outer reflection is $3.2\times10^{-8}$ ergs.cm$^{-2}$.sec$^{-1}$ which is 0.64 times of the flux in power-law component. This fact indicates that the flux reaching to the observer from the corona is greater than that reaching to the disk. Another very puzzling fact is that the flux in outer reflection component is four times the flux in inner reflection component. Stronger Compton hump for the outer reflection compared to inner reflection in many other fitting results also indicates larger flux in outer reflection component compared to inner reflection in energy range, 1 eV - 1 MeV (for example Fig. 5 in \citealt{Zdziarski2021MAXI} for epochs 1-4; Fig. 4 in \citealt{Marino2021} for epoch 2; Fig. 10 in \citealt{Liu2023} for source V404 Cyg). As all the released gravitational energy will be concentrated in the inner region due to proximity to the BH, it is natural to think that the power-law continuum that is the most dominant component of the spectra will also be concentrated in the inner region. However, with the centrally concentrated coronal radiation, it is difficult to imagine that the inner region of the disk gets only $\sim 1/10$th of the power-law flux\footnote{There might be some effect due to different geometry of the power law-emitting corona and the disk, but that would be on the level of $\sim$ 1.}. The simplest possibility for resolving the issue of very low inner reflection is to assume that the inner disk is being covered by the corona, which reduces the observed flux in the inner reflection component. 

\subsubsection{Fitting that includes the covering of the inner disk}
\label{sec_model_C1}
If the inner region of the disk is covered by the optically thick corona, we would expect that most of the inner reflection component is up-scattered by the corona and contributes to the power-law component. Thus, we introduced our model B.1 which is based on model B, but with the normalization of inner reflection reduced. For model B, the normalization is fixed according to the distance to the source; whereas for model B.1, the normalization is further reduced to 10\% of this value to mimic the covering of the inner disk by the corona (10\% is chosen as it is approximately the ratio of the observed power-law flux and inner reflection flux in model B). The best-fit parameters are given in the last column of Table \ref{table_fit_results}. We were able to find an equally good fit with model B.1, but with ten times higher luminosity and ionization of inner reflection compared to model B. It indicates the crucial fact that the fitting of inner reflection is not driven by a relativistic iron line, but by its contribution to the continuum (as seen in Fig. \ref{fig_bestfit_kynrefionx}). This is due to already very high inner disk ionization in model B. By increasing the ionization further, the inner reflection in model B.1 displays a similar spectral shape, but higher normalization. 
The goodness of the fit with model B.1 indicates that the inner disk covered with an optically thick corona can offer a plausible geometry for the observation analyzed in this work. Although, this geometry indicates that the disk is extended close to the BH, the obscuration of a major part of the inner disk by corona resembles a truncated disk geometry.

\subsubsection{Comptonization of inner reflection using MONK}
\label{sec_monk_results}
To check the possibility of reduction of inner reflection by its Comptonization in the slab corona above the disk, we used the relativistic Monte Carlo radiative transfer code MONK \citep{Zhang2019}. In the hard state, we did not see any disk blackbody emission due to intrinsic dissipation. Therefore, we assumed only the Comptonization of the inner reflection component here, which also includes the reprocessed blackbody. To keep things simple, we restricted our approach to the Newtonian regime and provided the unabsorbed inner reflection component of model B.1 (Section \ref{sec_model_C1}) as the seed spectrum for Comptonization. We assumed that the disk is emitting the seed spectrum as a cosine source and the thin slab of hot electrons as corona is present just above the disk. The electron temperature and optical depth of the corona are set according to the requirement of two factors: (i) the observed power law and (ii) the ratio of observed flux in inner reflection and power-law component in model B.1, which is 0.1 at 66$^\circ$ inclination of the source. The unscattered flux that leaks through the corona becomes the observed inner reflection component. The result of this experiment is presented in Fig. \ref{fig_monk_with_modelC1}(A), as broadband spectral output from MONK. To achieve the observed constraints, we require the optical depth of the corona to be 0.282 and temperature of the electrons to be 160 keV. The integrated term is used to denote the total spectrum integrated over 2$\pi$ solid angle (upper hemisphere above the disk or lower hemisphere below the slab corona). Otherwise, the spectrum at an inclination of 66$^\circ$ is indicated specifically. The components of unabsorbed model B.1 (by setting N$_{\rm H}=0$) are overplotted with corresponding MONK outputs in Fig. \ref{fig_monk_with_modelC1}(B). It shows that the above-mentioned coronal properties reproduce the observed power law quite accurately. Although there is some change in the shape of the unscattered spectrum compared to the incident spectrum, it appears that the covering of the disk by such corona can credibly lead to the required reduction in the observed inner reflection flux. 

However, although we would be able to achieve the observed power law and inner reflection with the slab corona lying above the disk, it is crucial to check whether this configuration can stand as a possible steady-state solution or not. In the MONK simulation, we started with the seed spectrum emitted by the disk for Comptonization (pink dashed line in Fig. \ref{fig_monk_with_modelC1} (A)). As a result, not only we get the observed power law, observed unscattered inner reflection, but also a Comptonized spectrum illuminating the disk (yellow dashed line in Fig. \ref{fig_monk_with_modelC1} (A)). This illumination will be reprocessed by the disk and emitted as the seed spectrum for Comptonization in the next step. A steady-state solution demands that the total integrated flux (in energy band 1 eV - 1 MeV) of the Comptonized power law illuminating the disk should be the same as the seed spectrum for Comptonization; however, this is not the case here. The Comptonized power law illuminating the disk integrated over 2$\pi$ solid angle ($3.69\times10^{-7}$ ergs.cm$^{-2}$.sec$^{-1}$) has 2.75 times more energy flux compared to the integrated seed spectrum ($1.34\times10^{-7}$ ergs.cm$^{-2}$.sec$^{-1}$), also denoted in the top left corner of Fig. \ref{fig_monk_with_modelC1} (A). Thus our model predicts more illumination, thus, more seed photons than we originally used to produce the power law. This indicates that static slab corona lying above the disk can not be a steady-state equilibrium solution to explain the required reduction of inner reflection component and some kind of anisotropy in the coronal emission may solve the problem as discussed in next section. To find the effect of Comptonization by different coronal geometries for spectra in hard state, we also refer to \cite{Poutanen2018}.

\subsubsection{Possibilities for required anisotropy}
The discrepancy between the Comptonized flux illuminating the disk and the seed spectrum required for Comptonization (described in Section \ref{sec_monk_results}) could, in principle, be overcome by introducing some anisotropy in the emission of the corona. One such possibility would be outflowing corona which is in line with the observed jet for this source \citep{Bright2020}. Following the procedure from Section 4.8 of \cite{Rybicki1979}, we find that moving corona with velocity $\sim0.2-0.3c$ ($c$ is the speed of light) perpendicular to the disk will lead to the integrated seed flux received by the corona from the disk reduced by $\sim2-3$ relative to the flux emitted by the corona toward the disk. From MONK simulation with outflowing slab corona covering the disk, we also find $0.3c$ velocity is sufficient to make the model as a steady-state solution for this observation. \cite{Beloborodov1999} carried out a similar analysis of bringing anisotropy in the coronal emission through its bulk motion to explain the observed power law and low reflection fraction in the hard state of Cyg X-1, which resulted in a velocity of $\sim0.3c$. Recent observations of polarization from Cyg X-1 even requires the slab corona to be moving at $\sim0.6 c$ \citep{Poutanen2023}. If we did not consider any obscuration of the disk by the corona like in a lamp-post geometry, the velocity of the corona would have to be $\sim0.7-0.8c$ to give the observed fraction of inner reflection flux relative to power law at 66$^\circ$ inclination (0.13 for model B, 0.1 for model B.1).

\subsubsection{Past analyses and comparison with current work}
As our primary objective here is reflection, we fit the RS considering its shape and flux, and we kept its normalization disconnected from the power law to avoid the uncertainty in disk illumination by the corona due to its uncertain geometry and/or velocity. It is only after we find the best fit, that we can discuss whether it can be interpreted with some corona geometry. In a similar line, there are many works that have fitted the observed data from the same source with arbitrary normalization of the power law and reflection components; even without worrying about the consistency of reflection from its shape and flux. Previous analyses of MAXI J1820+070 through its different states of the outburst were studied by \cite{Buisson2019, Chakraborty2020}, with two-component corona in lamp-post geometry illuminating the inner and outer disk. They have also found that an accretion disk may reach very close to the BH. Although their outer reflection remain lower compared to inner reflection as expected with centrally concentrated corona, their inner reflection is still too low with respect to the power-law component (first panel, Fig. 6 in \citealt{Buisson2019}\footnote{In \citealt{Buisson2019} it is mentioned that self-consistent reflection fraction is incorporated. However, as shown in the work of \citealt{Dauser2016} describing the reflection fraction in \textit{RELXILLLPCP} model, self-consistency should lead to larger reflection compared to illuminating power law for a low coronal height, instead of very low inner reflection found in Fig. 6 of \citealt{Buisson2019}}), similar to our result. This is impossible to achieve with static lamp-post corona above the disk.

On the other hand, \cite{Zdziarski2021MAXI}, who also considered two power-law corona components, reached the conclusion that the disk had to be truncated. In their work, they fit the same data with more acceptable values of inclination (the same as used in our current work) and iron abundance. The recent broadband study of \cite{Banerjee2024} also offered the same conclusion. The physical model of a jet emitting disk (JED) acting as the corona \citep{Marino2021} fit the data equally well with truncated disk geometry. In this scenario, hot advective accretion flow or JED serves as the corona, which lies in the same plane of the disk. The solid angle subtended by the disk to the corona is smaller compared to the 2$\pi$ solid angle of the upper hemisphere, resulting in a lower observed flux in the reflection component than in the power-law component. However, to achieve more flux in outer than inner reflection component, some obscuration of the inner disk region and/or flaring of the outer disk is still needed as suggested in Fig. 4 of \cite{Zdziarski2021MAXI} or in Fig. 10 of \cite{Marino2021}.

The observed timing properties of this source in the low/hard state show large lags between the coronal emission and the disk reflection, which \cite{Kara2019} interpreted with an elongated corona (modeled as two component lamp-post corona) illuminating a disk reaching close to the BH. The same lags can be explained also by the illumination of the outer disk by the central corona in the truncated disk geometry (\citealt{DeMarco2021, Kawamura2022}). Although, we did not try to model the timing properties, from the spectral modeling, we propose a possible geometry of outflowing slab corona covering the disk. The inner region of the disk remains covered by the slab corona and, from the observed flux, we can see that the outer reflection is more dominant than the inner reflection. Therefore, the proposed geometry becomes very similar to the truncated disk geometry; as a consequence, this could, in principle, explain the observed lag between the power law and the reflection.

Although all the works mentioned above are able to fit the data component wise, none of them have fitted the reflection component by simultaneously considering its shape and flux, as done in this work. Additionally, the proper modeling of Comptonization and obscuration by the corona was only carried out only in few works to describe the observations, for instance, \cite{Steiner2017, Poutanen2018} (for AGNs, see \citealt{Petrucci2001, Malzac2005, Wilkins2015}). To investigate further about the geometry with appropriate modeling of Comptonization, we need to include more coronal geometries in the existing MONK code; for instance, outflowing slab corona or truncated disk geometry. The Comptonization of the reflection spectrum in the general relativistic framework also has to be further developed.

\subsection{Limitations of our study and possibilities for its extension}
\label{sec_caveats}
Naturally, there are a few limitations of our study that need to take into account to understand the robustness of the results. The atomic database of {\tt reflionx\_hd} table is not as rich as in {\tt XILLVER} tables. The fixed high energy cutoff value of {\tt reflionx\_hd} table used in this work also restricts us to a narrow parameter regime. In addition, there can be two different Comptonizing components giving rise to two different power law, as investigated in \cite{Zdziarski2021MAXI}. The self-consistent normalization of the RS from both the flux and the shape is possible only because of the well known distance and inclination of the source. We do not have this freedom for most of the observed XRBs. To study the geometry, we only considered the Comptonization of the inner reflection emitted by the disk. A self-consistent modeling of coupled disk-corona system with possible energy exchange between them is beyond the scope of this work. The incorporation of intrinsic viscous dissipation in the fitting also demands development of new reflection table. Above all, we have analyzed only one spectra in low/hard state of one source in this study. In addition to addressing the difficulties presented in this work to explain the spectral flux distribution in fitted spectral components, we also need to study the observed timing properties to understand the true geometry of the source.

Despite of the above limitations, it will be interesting to extend our study to other XRBs in the low/hard state and find the inner radius of the disk. If the source has well-known distance, inclination, and mass values for the BH (e.g. V404 Cyg, \citealt{MillerJones2009, Khargharia2010}), then the simultaneous fitting of shape and flux of RS should give very accurate value of inner radius of the disk. Even though the distance and inclination of GX 339-4 is uncertain (\citealt{Basak2016}), constraining them to some acceptable range and simultaneous fitting the flux and shape of the RS for multiple observations together can shed some light on the inner radius of the disk in low/hard state for that source. Due to repeated outbursts of this source it is observed and studied many times. There is a large scatter about the inferred inner radius of the disk (Fig. 11 in \citealt{Garcia2015}).


\section{Conclusions}
\label{sec_conclusions}
In this work, we have fit both the shape and the flux of the observed RS in the low/hard state of the XRB MAXI J1820+070. We have examined whether we can exclude the possibility of disk extending close to the BH from the perspective of reprocessed black body due to disk illumination. From the flux consideration we have attempted to constrain the geometry of the source. The key findings are summarized below.
\begin{itemize}
    \item Consistently carrying out the fitting, with both the shape and the flux of the observed RS taken into consideration, gives the exact estimate of the density and the ionization of the disk. The consistent best fit with one power-law component suggests that the disk is extended very close to the BH in the low/hard state of XRB. The fitting of inner reflection component gives quite high disk ionization; thus, it is primarily driven by reflection continuum shape and its contribution to the total flux rather than by the shape of the relativistic iron line.
    \item The density and ionization of the disk are high enough to give rise to the observed reflected flux and low enough to make the reprocessed blackbody negligible. It is not possible to exclude the possibility for disk extending very close to the BH in low/hard state from the perspective of reprocessed blackbody.
    \item The total flux in reflection is smaller than the flux in the power-law component. At the same time, the inner reflection is much lower than the outer one, which is impossible for a centrally located corona. We show that covering of the inner disk by static slab corona may lead to the necessary level of the power-law component as a result of the Comptonization of re-processed emission and, at the same time, to the apt level of transmitted unscattered inner reflection. On the other hand, the flux in generated power-law component illuminating the disk is higher than the assumed seed photon flux for Comptonization.
    \item An outflowing corona with some bulk velocity ($\sim 0.2-0.3 c$) could, in principle, reduce the illumination of the disk and lead to the same flux in the seed spectrum seen for Comptonization and in a Comptonized power law emitted by the corona toward the disk. This has to be investigated thoroughly with the appropriate Comptonization in the outflowing corona.
    

\end{itemize}

\begin{acknowledgements}
SRD and MD thank Andrzej A. Zdziarski and Chris Done for many stimulating scientific discussions. SRD also thanks Peter Boorman, Ji\v{r}\'{i} Svoboda and Craig A Gordon for their help in learning {\tt XSPEC}. The authors also thank the reviewer for many critical constructive comments which make the conclusion of the manuscript more robust. SRD, MD, MB, JH and VK thank GACR project 21-06825X for the support, as well as the institutional support from RVO:67985815. WZ  acknowledges NSFC grant 12333004, and support by the Strategic Priority Research Program of the Chinese Academy of Sciences, grant no. XDB0550200.
\end{acknowledgements}

\bibliographystyle{aa} 
\bibliography{ref} 

\begin{appendix}
\section{Fitting shape and flux simultaneously through iteration of density}
\label{appendix_density_iteration}

\subsection{The model}
\label{appendix_model_den_ite}
Besides model B (presented in section \ref{sec_kyn_model_and_fit}), here we build another alternative model with which also we can fit simultaneously the shape and flux of the RS (reflection spectra). Here reflection is modeled by ({\tt relconv}$\otimes${\tt reflionx\_hd}) as done in model A (Section \ref{sec_problem}). Fitting the RS using the model ({\tt relconv}$\otimes${\tt reflionx\_hd}) gives values of $\xi$, $n$ and $r_{\rm in}$ of the disk. {\tt reflionx\_hd} model in {\tt XSPEC} has normalization parameter which we keep free while fitting. First we fit the spectra keeping the density parameter free which essentially fits only the shape of the spectra because using arbitrary density and normalization {\tt XSPEC} rescale the model spectrum to the desired flux value. For fixed values of parameters (e.g. photon index, $\xi$, $n$) of {\tt reflionx\_hd} table, it gives a specific reflection spectrum. Now when we apply the {\tt relconv} model on {\tt reflionx\_hd} table, it does the following: (i) takes that specific spectrum depending on the input parameters of the {\tt reflionx\_hd} table, (ii) adds contribution from each radius according to $r^{-q}$, i.e., emissivity index parameter of {\tt relconv} model, (iii) integrates these radial flux contributions over the disk area from $r_{\rm in}$ to $r_{\rm out}$ after incorporating all the relativistic effects. Therefore, in {\tt XSPEC}, this final model spectrum is used to fit the observed reflection spectrum whenever we use ({\tt relconv}$\otimes${\tt reflionx\_hd}) to model the reflection. Note that although the {\tt relconv} model takes radial flux contribution according to emissivity profile, it takes only one single spectrum from {\tt reflionx\_hd} table corresponding to one $\xi$ value for all radii. So, the fitted value of $\xi$ gives an average value of $\xi(r)$ which best fits the shape and we term it as $\xi_{\rm shape}$.

From the point of view of the flux, the assumed radial emissivity profile (Equation \ref{eqn_emi_profile}) demands the same radial profile of $\xi(r)$ as we are not considering any radial variation of density,
\begin{equation}
    \xi(r)=\xi_1r^{-q},
\end{equation}
where $\xi_{\rm 1}$ is the ionization at $r=1$ and $r$ is always in units of $R_{\rm g}$. Plugging the radial profile of $\xi$ in Equation \ref{eqn_F_refl}: 
\begin{equation}
    F_{\rm refl}(r)=\frac{n\xi(r)}{4\pi}=\frac{n\xi_1}{4\pi}r^{-q}.
\end{equation}
It is important to remember that the above relation holds true because the slab is in radiative equilibrium with the incident radiation which makes the total reflected flux in the energy range 1 eV - 1 MeV equal to the incident flux. As $\xi_{\rm shape}$ represents some average value of $\xi(r)$ fitting the shape of the spectra, to make a comparison with it, we must find some average value of $\xi(r)$ corresponding to flux, namely, $\xi_{\rm flux}$. To find that, we take flux weighted average of $\xi$ over the disk area, which gives
\begin{multline}
\label{eqn_xi_flux}
    \xi_{\rm flux}=\frac{\int_{r_{\rm in}}^{r_{\rm out}}\xi(r)F_{\rm refl}(r)2\pi rdr}{\int_{r_{\rm in}}^{r_{\rm out}}F_{\rm refl}(r)2\pi rdr}
    =\frac{\int_{r_{\rm in}}^{r_{\rm out}}\xi_{\rm 1}r^{-q}\left(\frac{n\xi_{\rm 1}}{4\pi}\right)r^{-q}2\pi rdr}{\int_{r_{\rm in}}^{r_{\rm out}}\left(\frac{n\xi_{\rm 1}}{4\pi}\right)r^{-q}2\pi rdr}\\
    =\xi_{\rm 1}\left(\frac{2-q}{2-2q}\right)\left(\frac{r^{2-2q}_{\rm out}-r^{2-2q}_{\rm in}}{r^{2-q}_{\rm out}-r^{2-q}_{\rm in}}\right).
\end{multline}
The total radiation emitted by one side of the disk is
\begin{multline}
\label{eqn_L_re}
    L_{\rm refl}=\int_{r_{\rm in}}^{r_{\rm out}}F_{\rm refl}(r)2\pi rdr\times R^2_{\rm g}
    =\frac{\xi_{\rm 1}n}{2(2-q)}\left(r^{2-q}_{\rm out}-r^{2-q}_{\rm in}\right)\times R^2_{\rm g}\\
    =\xi_{\rm flux}n\frac{(1-q)}{(2-q)^2}\frac{\left(r^{2-q}_{\rm out}-r^{2-q}_{\rm in}\right)^2}{\left(r^{2-2q}_{\rm out}-r^{2-2q}_{\rm in}\right)}\times R^2_{\rm g}.
\end{multline}
At inclination $\theta$ and distance $D$ from the source, the observed reflected flux will be 
\begin{equation}
\label{eqn_F_re_o}
    F_{\rm refl,o}=L_{\rm refl}\times \left(\frac{\cos\theta}{\pi D^2}\right).
\end{equation}
For the limiting case of $r_{\rm out}\gg r_{\rm in}$ and $q=3$, $\xi_{\rm flux}=\xi_{\rm 1}r^{-3}_{\rm in}/4$, $L_{\rm refl}=2\xi_{\rm flux}n(r_{\rm in}R_{\rm g})^2$, leading to 
\begin{equation}
\label{eqn_F_re_o_limit}
    F_{\rm refl,o}=2\xi_{\rm flux}n(r_{\rm in}R_{\rm g})^2\times(\cos\theta/\pi D^2).
\end{equation}

From the fitting of observed spectra, we find the unabsorbed reflected flux within the energy range 1 eV - 1 MeV using the {\tt flux} command in {\tt XSPEC}, which gives true $F_{\rm refl,o}$. This large energy range covers essentially all possible radiation emitted by the slab. Note that this becomes possible only because the energy grid of the {\tt reflionx\_hd} table covers this large range. Using Equation (\ref{eqn_L_re}) \& (\ref{eqn_F_re_o}), from $F_{\rm refl,o}$ we find $\xi_{\rm flux}$ which should match with $\xi_{\rm shape}$ from fitting. For $n=10^{15}$ cm$^{-3}$, always $\xi_{\rm flux}$ turns out to be larger than $\xi_{\rm shape}$. From Equation (\ref{eqn_L_re}), it is clear that by increasing $n$, we can reduce $\xi_{\rm flux}$ and meet the demand of high flux while $\xi_{\rm flux}=\xi_{\rm shape}$. After fixing the $n$ corresponding to $\xi_{\rm flux}=\xi_{\rm shape}$, we again fit the data since $n$ changes the spectral shape.  $\xi_{\rm shape}$ and $r_{\rm in}$ are modified due to fitting and again new $n$ is required to converge. We iterate until a fit is achieved when $\xi_{\rm shape}$ become within the 5\% range of $\xi_{\rm flux}$. Therefore, model C for our fitting becomes
\newline
model C = {\tt TBabs}$\times$({\tt cutoffpl} + {\tt relconv}$\otimes${\tt reflionx\_hd} ($\xi_{\rm flux}=\xi_{\rm shape}$)+ {\tt relconv}$\otimes${\tt reflionx\_hd} ($\xi_{\rm flux}=\xi_{\rm shape}$)).
\newline
The final fit that follows $\xi_{\rm flux}=\xi_{\rm shape}$ will possibly not be as good as when there is no restriction in density. Through iteration of density we are able to match $\xi_{\rm shape}$ and $\xi_{\rm flux}$, and thus fit the observation simultaneously by its shape and flux.

\begin{table}
	\centering
	\caption{Best-fit model parameters with consistent flux and shape for model C.}
	\label{table_appendix_den_ite}
	\begin{tabular}{ccc}
		\hline
		Component & Parameter & model C\\ 
		\hline
		Interstellar absorption & N$_{\rm H}$(10$^{22}$ cm$^{-2}$) & 0.14(f)\\
		\hline
		   & Photon Index & 1.400\\  
         Power law & cut-off(keV) & 300(f)\\
         & Norm & 0.678\\
         & Flux & 4.89$\times10^{-8}$\\
         \hline
          & $r_{\rm in}$ ($R_{\rm g}$) & 2.71\\
          & $r_{\rm out}$ ($R_{\rm g}$) & 1000(f)\\
          & $\xi_{\rm shape}$ ($\approxeq\xi_{\rm flux}$) & 8.84$\times10^{3}$\\
          Inner reflection & $\xi_{\rm in}$ & 3.53$\times10^{4}$\\
          & $n$ ($10^{15}$ cm$^{-3}$) & 1.49$\times10^{4}$\\
          & Norm & 1.24\\
          & Flux & 4.27$\times10^{-9}$\\
          \hline
          & $r_{\rm in}$ ($R_{\rm g}$) & 445\\
           & $r_{\rm out}$ ($R_{\rm g}$) & 1000(f)\\
           & $\xi_{\rm shape}$ ($\approxeq\xi_{\rm flux}$) & 100(f)\\
          Outer reflection & $\xi_{\rm in}$ & 231\\
          & $n$ ($10^{15}$ cm$^{-3}$) & 1.10$\times10^3$\\
          & Norm & 198.91\\
          & Flux & 3.08$\times10^{-8}$\\
          \hline
           & $\chi^2/\nu$ & 2141/1865\\
          \hline          
	\end{tabular}
\flushleft
\textit{Notes.} {Due to the delicacy involved in manual iteration, reasonably good fit (with $\xi_{\rm shape}$ and $\xi_{\rm flux}$ lie within 5\% range) is achieved, not the best fit, which also prohibits to find the errors associated to the fitted parameters. $\xi$ at inner radius of the disk ($\xi_{\rm in}$) is computed from the fitted $\xi_{\rm shape}\approxeq\xi_{\rm flux}$. The tabulated flux is the unabsorbed one in units of ergs.cm$^{-2}$.sec$^{-1}$, calculated from {\tt XSPEC} within energy range 1 eV - 1 MeV. The frozen parameters during fitting are indicated as (f).}
\end{table}

\subsection{Results}
\label{appendix_results_den_ite}
\begin{figure}
\centering\includegraphics[width=\columnwidth]{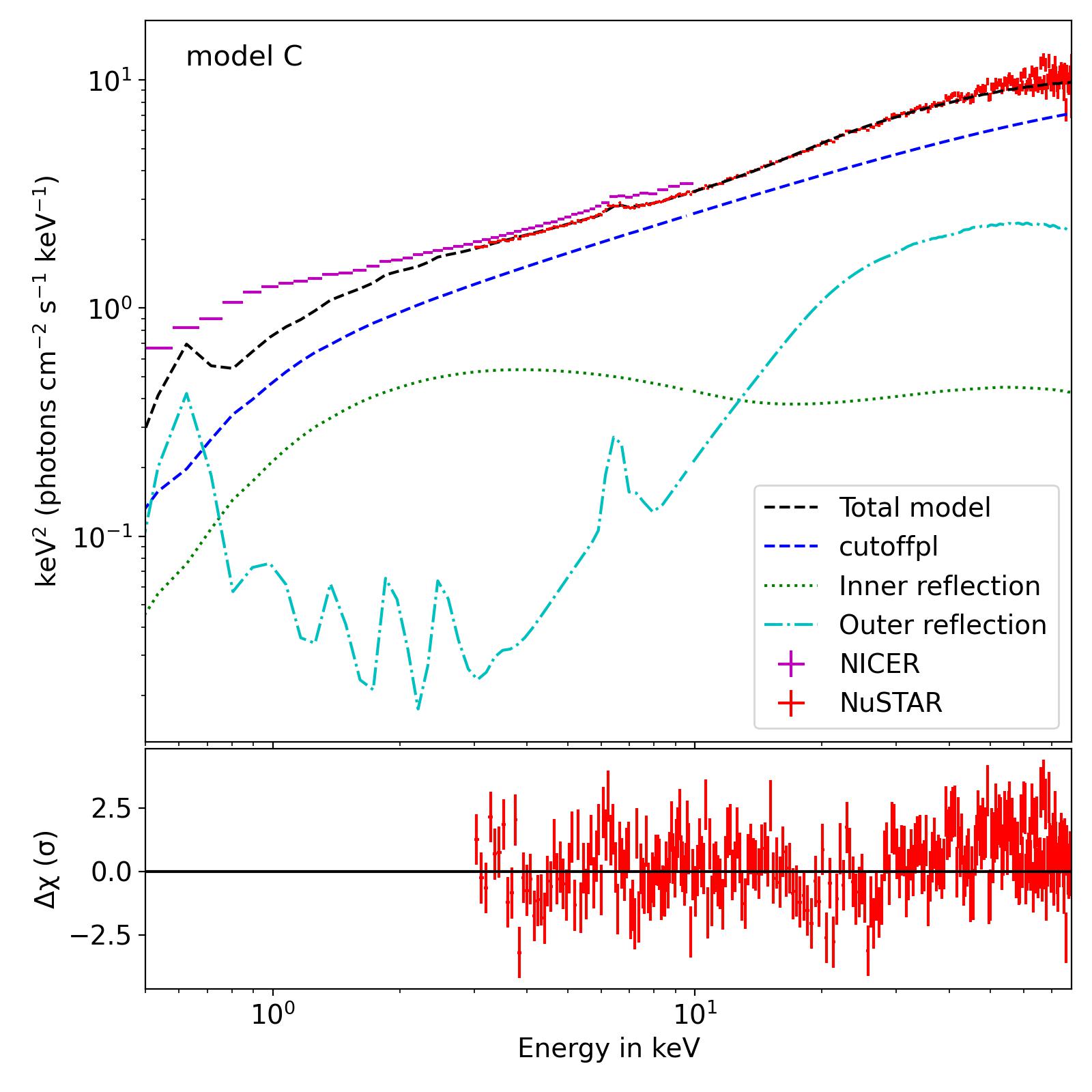}
\caption{Reasonably good-fit model C for {\it NuSTAR} FPMA observation along with its components when reflection $\equiv$ {\tt relconv}$\otimes${\tt reflionx\_hd}. Densities for reflection components are iterated to make the fitting consistent from shape as well as flux. The fitted model parameters are presented in Table \ref{table_appendix_den_ite}. Simultaneous {\it NICER} observation is also overplotted in the energy range of 0.5-10.0 keV which is sufficient to understand the consistency or inconsistency of reprocessed blackbody in soft X-ray. The best fitted model components turn out to be very similar with the results of model B which is shown in Fig. \ref{fig_bestfit_kynrefionx}.}
\label{fig_bestfit_relconv}
\end{figure}

For the fitting with model C, densities for both the reflection components were first kept free and the best fit was achieved. Progressing toward consistency, we varied the density and fitted iteratively, as described at the end of section \ref{appendix_model_den_ite}. We fixed the ionization to the most neutral case provided by the tables, $\xi=100$ for the outer reflection component. The reasonable good fit model is shown in the Fig. \ref{fig_bestfit_relconv}, and the relevant parameters are presented in Table \ref{table_appendix_den_ite}. The statistics of the consistent fit is $\chi^2/d.o.f=2141/1865$, slightly worse than the best fit when densities were kept free ($\chi^2/d.o.f=2123/1865$). The density to make $\xi_{\rm flux}=\xi_{\rm shape}$ for inner and outer reflection becomes $1.5\times10^{19}$ cm$^{-3}$ and $1.1\times10^{18}$ cm$^{-3}$, respectively. The iteration in model C is delicate due to the strong dependence of flux on $r_{\rm in}$, $n$, and $\xi$. Therefore, once we achieve a reasonably good fit, we avoid finding the best fit. Consequently, we cannot estimate the errors in the model parameters using {\tt XSPEC} because that is possible only after achieving the best fit. That is the reason why no error for parameters of model C is presented in Table \ref{table_appendix_den_ite}. With the consistent parameters, the theoretically calculated flux remains within the 5\% range of unabsorbed flux estimated through {\tt XSPEC} for the energy range 1 eV - 1 MeV (based on the criterion we follow for convergence after iterations as mentioned in Section \ref{appendix_model_den_ite}). This confirms that the fit is almost consistent simultaneously with the shape and flux of the observed RS. The fitting results are presented in Table \ref{table_appendix_den_ite} and in Fig. \ref{fig_bestfit_relconv} which seem very similar to the results of more accurate model B presented in Section \ref{sec_kyn_model_and_fit}.

\section{Fitting {\it NuSTAR} and {\it NICER} together for MAXI J1820+070}
\label{appendix_fit_nicer_nustar}
Here we present the fitting of quasi-simultaneous data of {\it NuSTAR} FPMA, {\it NuSTAR} FPMB and {\it NICER} with self-consistent normalization considering both the shape and the flux of the RS. The energy range for fitting {\it NuSTAR} data is 3.0-78.0 keV whereas for {\it NICER} we fit 0.5-10.0 keV. One such model with self-consistent normalization of RS is model B (Section \ref{sec_kynrefionx_model}). Fitting of only {\it NuSTAR} FPMA data with model B is presented in Fig. \ref{fig_bestfit_kynrefionx} and Table \ref{table_fit_results} (Section \ref{sec_kynrefionx_result}). To fit the data from different instruments, we need to incorporate a constant factor for cross-normalization. In addition, following previous works \citep{Wang2020, Wang2021, Ludlam2021, Ludlam2022, Kumar2022, Feng2022, Moutard2023, Svoboda2024ApJ, Svoboda2024ApJSwiftSoft}, to handle the different calibration features present in the low energy data from {\it NICER}, we incorporate a Gaussian absorption line ({\tt gabs}) and absorption edge ({\tt edge}) to fit the line feature below 1 keV and edge around 2.4 keV. These models ({\tt gabs} and {\tt edge}) are applied only to fit the {\it NICER} data. This makes model B.2 for fitting {\it NuSTAR} + {\it NICER} data together. 
\newline
model B.2 = {\tt constant}$\times${\tt gabs}$\times${\tt edge}$\times$model B = {\tt constant} $\times$ {\tt gabs} $\times$ {\tt edge} $\times$ {\tt TBabs} $\times$ ({\tt cutoffpl} + {\tt kynrefionx}({\tt reflionx\_hd}) + {\tt kynrefionx}({\tt reflionx\_hd})).

\begin{figure}
\centering\includegraphics[width=\columnwidth]{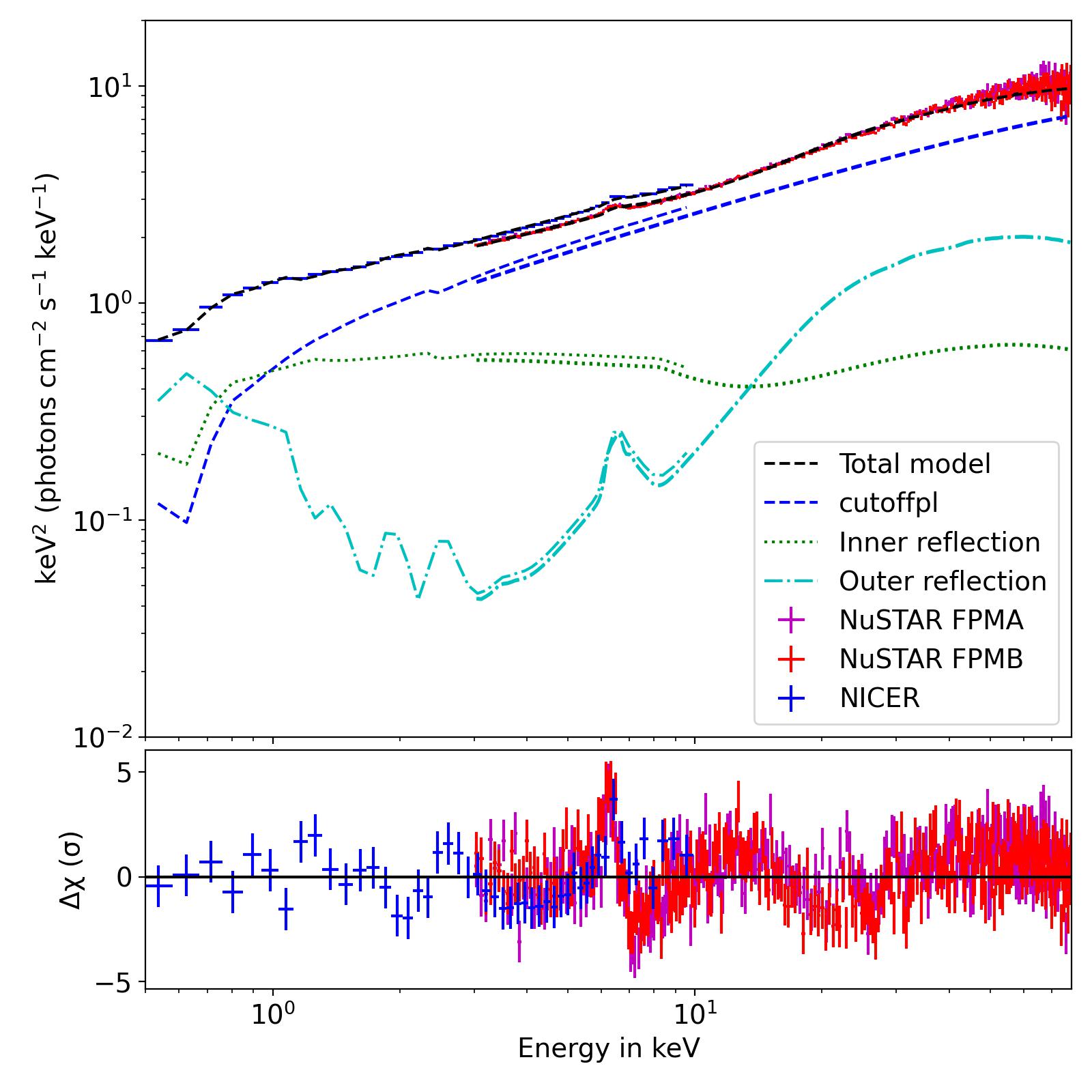}
\caption{Spectral fit for {\it NuSTAR} FPMA (magenta), {\it NuSTAR} FPMB (red) and {\it NICER}  (blue) with model B.2. Top panel shows the unfolded spectra and bottom panel shows the data residuals. Although residuals indicate that this is not a perfectly fitted iron line, the reprocessed blackbody remain sufficient and consistent to fit the data from {\it NICER} along with {\it NuSTAR}. The fitted model parameters are presented in Table \ref{table_fit_nicer_nustar}. Data is rebinned using the command ``setplot rebin 100 10'' to present in the plot.}
\label{fig_fit_nuAnuBni}
\end{figure}

The relevant parameters of the best fitting are mentioned in Table \ref{table_fit_nicer_nustar}. Corresponding unfolded spectra and the residuals are plotted in top and bottom panel of Fig. \ref{fig_fit_nuAnuBni}. The best fitting leads to $\chi^2/\nu=4519/3904\sim1.16$ (for {\it NuSTAR} FPMA $\chi^2=2246$ for 1874 bins, for {\it NuSTAR} FPMB $\chi^2=2165$ for 1874 bins, and for {\it NICER} $\chi^2=108$ for 171 bins respectively). Note that the $\chi^2$ value for {\it NICER} becomes much less than the number of bins which is basically due to added 1.5\% systematic error while reducing the level 3 data following complete product pipeline of {\tt NICERDAS}. On the contrary, if we do not add {\tt gabs} or {\tt edge} component then the $\chi^2$ value becomes reasonable but residuals clearly indicate the presence of such components.  

Residuals in Fig. \ref{fig_fit_nuAnuBni} (bottom panel) indicate some feature still present around iron line emission. This is possibly due to some abrupt change in the iron line shape in {\tt reflionx\_hd} table with change in density or ionization. As an example, notice the change in iron line shape around 6.4 keV in the right panel of Fig. \ref{fig_reflionxhd_density_xi_variation} for $\xi$=278.3 with respect to $\xi$=100.0 and $\xi$=774.3. This kind of diminished iron line can be noticeable for some other values of parameters also. We verify this argument by including another Gaussian line with existing two reflection components in fitting which diminishes such feature present in residuals significantly. Coarser energy grid of the {\tt reflionx\_hd} table (energy spacing between two grid points is around 0.1 keV at 6.4 keV) may have some effect in fitting iron line. Additionally, the data we are fitting from {\it NICER} and {\it NuSTAR} are not exactly simultaneous. This non-simultaneity also can contribute to the residuals as the fitting is done assuming same model parameters for the observations from two instruments. Despite the fact that the iron line is not fitted perfectly well for simultaneous data from {\it NICER} and {\it NuSTAR}, reprocessed blackbody from reflection still become sufficient to fit the low energy data and remain consistent with the non observation of any other blackbody component in the hard state. Compared to fitting only the {\it NuSTAR} FPMA data (Table \ref{table_fit_results}, model B), primarily the density ($n$) and ionization ($\xi_{\rm in}$) of the inner reflection component change to fit the spectra of {\it NICER} in low energy along with {\it NuSTAR}. There is also slight change in ionization and density of outer reflection. The flux in different components remain almost same. This result denotes that density can also play a crucial role in fitting the broadband RS, along with the ionization parameter.

\begin{table}
	\centering
	\caption{Best-fit model parameters for model B.2.}
	\label{table_fit_nicer_nustar}
	\begin{tabular}{ccc}
		\hline
		Component & Parameter & Model B.2\\ 
		\hline
         & {\it NuSTAR} FPMA & 1.0(f)\\
         Constant & {\it NuSTAR} FPMB & 0.990$_{-0.001}^{+0.001}$\\
          & {\it NICER} & 1.092$_{-0.002}^{+0.002}$\\
		\hline
         Gaussian absorption & Line E (keV) & 0.614$_{-0.001}^{+0.001}$\\
         ({\it NICER}) & Width ($\sigma$) (keV) & 0.059$_{-0.002}^{+0.002}$\\
         & Strength (keV) & 0.124$_{-0.004}^{+0.004}$\\
          \hline
        Edge & edge E (keV) & 2.40$_{-0.03}^{+0.03}$\\
        ({\it NICER}) & Max Tau & 0.072$_{-0.007}^{+0.007}$\\
        \hline
		Interstellar absorption & N$_{\rm H}$(10$^{22}$ cm$^{-2}$) & 0.14(f)\\
        \hline
		   & Photon Index & 1.385$_{-0.007}^{+0.008}$\\  
         Power law & cut-off(keV) & 300(f)\\
         & Norm & 0.653$_{-0.013}^{+0.014}$\\
         & Flux & 5.00$\times10^{-8}$\\
         \hline
          & $r_{\rm in}$ ($R_{\rm g}$) & 1.43$_{-0.16}^{+0.27}$\\
          & $r_{\rm out}$ ($R_{\rm g}$) & 1000(f)\\
          & $L$ (L$_{\rm Edd}$) & 3.70$_{-0.41}^{+0.58}\times10^{-4}$\\
          Inner reflection & $\xi_{\rm in}$ & 6.35$\times10^{4}$\\
          & $n$ ($10^{15}$ cm$^{-3}$) & 5.92$_{-0.48}^{+0.46}\times10^{4}$\\
          & Norm & 1.11$\times10^5$(f)\\
          & Flux & 6.49$\times10^{-9}$\\
          \hline
          & $r_{\rm in}$ ($R_{\rm g}$) & 285$_{-30}^{+22}$\\
           & $r_{\rm out}$ ($R_{\rm g}$) & 1000(f)\\
           & $L$ (L$_{\rm Edd}$) & 1.09$_{-0.05}^{+0.05}\times10^{-3}$\\
          Outer reflection & $\xi_{\rm in}$& 388\\
          & $n$ ($10^{15}$ cm$^{-3}$) & 1.00$_{-0.11}^{+0.19}\times10^3$\\
          & Norm & 1.11$\times10^5$(f)\\
          & Flux & 2.42$\times10^{-8}$\\
          \hline
           & $\chi^2/\nu$ & 4519/3904\\
          \hline          
	\end{tabular}
\flushleft
\textit{Notes.} {Model B.2: i.e., model B with added Gaussian absorption, edge (for {\it NICER}), and a constant factor for cross-normalization across different instruments. The given errors for model B.2 represent 1 $\sigma$ range of fitted parameters. Luminosity ($L$) is a free parameter from which $\xi(r)$ is computed inside the model, and the value of which at inner radius ($\xi_{\rm in}$) is presented. The tabulated flux is the unabsorbed one in units of ergs.cm$^{-2}$.sec$^{-1}$, calculated from {\tt XSPEC} within energy range 1 eV - 1 MeV. The frozen parameters during fitting are indicated as (f).}
\end{table}

\end{appendix}

\end{document}